\documentclass[11pt,journal,twocolumn]{IEEEtran}

\usepackage{algorithm}
\usepackage{algorithmic}
\usepackage{amssymb}
\usepackage{cite}
\usepackage{graphicx}
\usepackage{amsmath}
\usepackage{times}
\usepackage{latexsym}
\usepackage{graphicx}
\usepackage{bm}
\usepackage{amssymb}
\usepackage[center]{caption2}
\usepackage{stfloats}
\usepackage{cases}
\usepackage{subfigure}
\usepackage{array}
\usepackage{setspace}
\usepackage{fancyhdr}
\usepackage{colortbl}
\usepackage{CJK}
\usepackage{tikz}
\usepackage{float}
\usepackage{color}
\usepackage{bm}
\usepackage{mathrsfs}
\usepackage{diagbox}
\usepackage{cuted}

\begin{document}
	\title{Robust Design for Intelligent Reflecting Surfaces Assisted MISO Systems}
	\author{Jiezhi Zhang, Yu Zhang, Caijun Zhong, and Zhaoyang Zhang}
	\maketitle

		\begin{abstract}
			In this work, we study the statistically robust beamforming design for an intelligent reflecting surfaces (IRS) assisted multiple-input single-output (MISO) wireless system under imperfect channel state information (CSI), where the channel estimation errors are assumed to be additive Gaussian.  We aim at jointly optimizing the transmit/receive beamformers and IRS phase shifts to minimize the average mean squared error (MSE) at the user. In particular, to tackle the non-convex optimization problem, an efficient algorithm is developed by capitalizing on alternating optimization and majorization-minimization techniques. Simulation results show that the proposed scheme achieves robust MSE performance in the presence of CSI error, and substantially outperforms conventional non-robust methods.
		\end{abstract}
		
	

%
%
%






\section{Introduction}
Intelligent reflecting surface (IRS) has recently emerged as a promising candidate to enhance the spectral and energy efficiency of future wireless communication systems \cite{Shi1,Shi2,Zhong1,Zhong0,Zhong2}.
Specifically, an IRS is composed of a large amount of  low-cost reflecting elements, each being able to passively  reflect the incident signal with a reconfigurable phase shift.
By smartly tuning the phase shifts, also known as passive beamforming, the signal reflected by the IRS can be adjusted towards the desired spatial direction.

In an IRS-assisted communication system, active precoding at the access point (AP) and passive beamforming at the IRS can be jointly designed to improve the system performance. In \cite{Wu2018IntelligentRS}, a joint active and passive beamforming design maximizing the total received signal power at the user was developed via the semidefinite relaxation (SDR) method. A novel discrete reflect beamforming design was investigated in \cite{Wu2018BeamformingOF} to minimize the transmit power at the AP. The authors of \cite{Yu2019EnablingSW} considered a secure wireless system with one legitimate user and one eavesdropper, where the secrecy rate was maximized based on the block coordinate descent (BCD) method, \textcolor{black}{while the impact of artificial noise on the secrecy beamforming design was studied in \cite{guan2020intelligent}.}


\textcolor{black}{However, all these prior works make the same assumption that perfect channel state information (CSI) is available, which is highly unlikely due to the lack of radio resources at the IRS. In general, the IRS-related channels can be separately estimated using the bilinear alternating least squares algorithm \cite{CEIRS}, whereas the direct channel can be estimated via traditional pilot-based approach.
Unfortunately, due to channel estimation error and the feedback latency, only imperfect CSI can be obtained in the practice wireless systems.} 
Motivated by this, we consider an IRS-assisted downlink multiple-input-single-output (MISO) system with a single user.
We address the problem of robust beamforming design with imperfect CSI. Specifically, assuming that the channel estimation error follows the complex Gaussian distribution,  we joint optimize active precoder at the AP, passive beamforming at the IRS, and one-tap equalizer at the user to minimize the average  mean squared error (MSE).

To tackle the non-convex optimization problem, we propose an alternating optimization (AO) algorithm based on the majorization-minimization (MM) technique \cite{Hunter2004ATO}. Closed-form solutions are obtained for the optimization variables during each iteration, which greatly reduce the computational complexity of the algorithm. In addition, the convergence of the proposed AO algorithm is established. Simulation results are presented to illustrate the performance of the proposed algorithm, and it is shown that the proposed scheme yields substantial performance gain over conventional non-robust design schemes.

\section{System Model And Problem Statement}
\subsection{System Model}
\begin{figure}[htbp]
	\centering
	\includegraphics[width=2in]{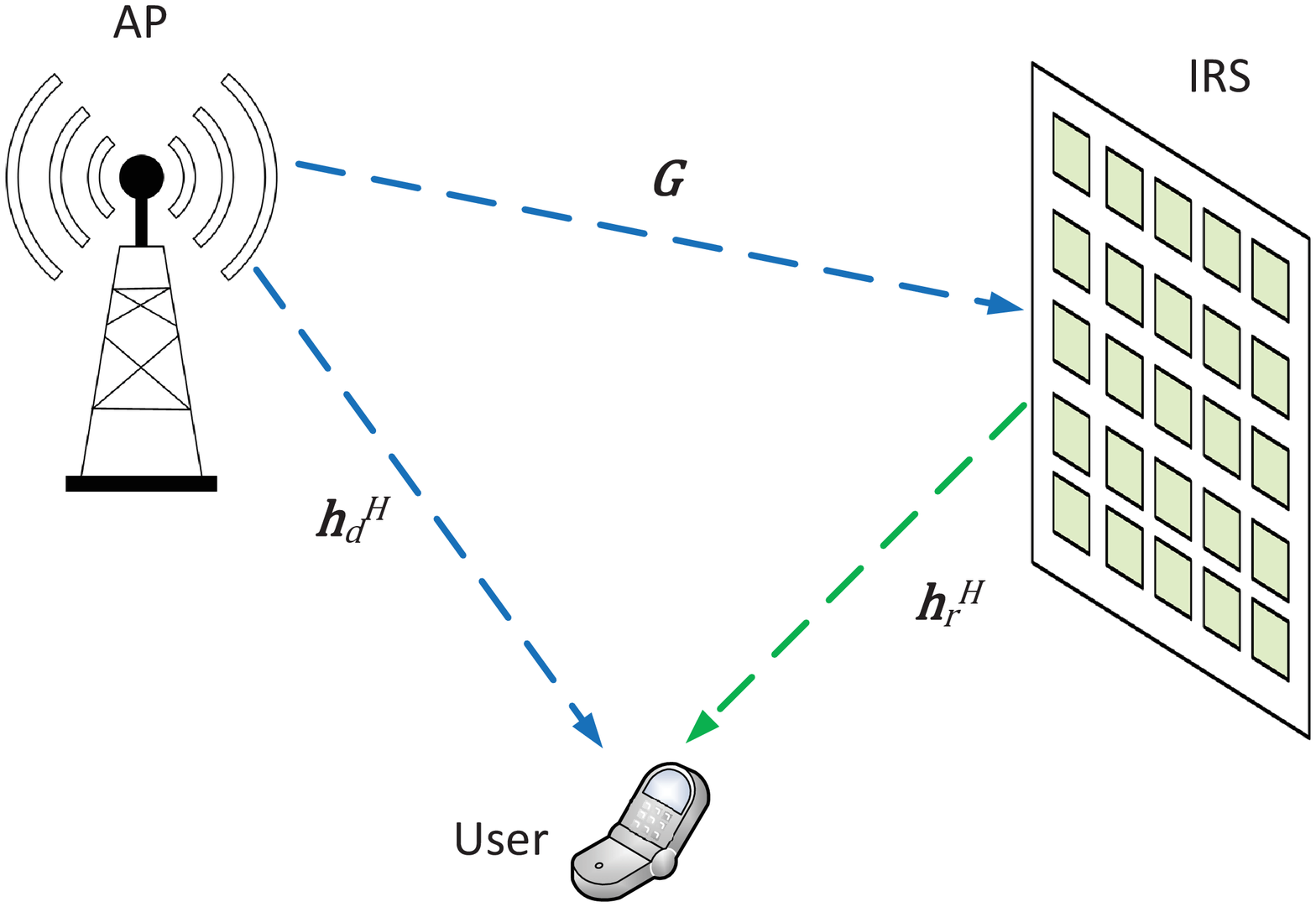}
	\caption{An IRS-aided MISO wireless system}
	\label{sys_model}
\end{figure}

In this section, we consider an IRS-assisted downlink MISO system consisting of one single-antenna user, one AP equipped with  $M$ transmit antennas, and one IRS with $N$ passive reflecting elements, as shown in Fig.\ref{sys_model}. 
\textcolor{black}{As in [6], the signals reflected by the IRS two or more times are ignored. Thus, the received signal at the user can be given by }
\begin{equation}
y = (\textbf h_r^H \boldsymbol \Theta \textbf G + \textbf{h}_d^H) \textbf w s + n_0,
\label{eq:downlink received signal r}
\end{equation}
where $\textbf{G} \in\mathbb C^{N \times M}, \textbf{h}_r\in \mathbb C^{N\times 1},$  $ \textbf{h}_d \in \mathbb C^{M\times 1}$ denote the AP-IRS link, IRS-User link, and AP-User link channels, respectively.  Also, $\boldsymbol{\Theta} \triangleq {\rm diag}(e^{j\theta_1}, \cdots, e^{j\theta_N})$ is the diagonal  reflection matrix of the IRS with $\theta_n \in [0,2\pi)$ being the corresponding phase shift.
Futhermore, $\textbf{w} \in \mathbb C^{M \times 1}$ represents the transmit beamformer at the AP satisfying $||\textbf{w}||^2 \leq P_0$, where $P_0$ is the maximum transmit power, while $s$ denotes the zero-mean complex Gaussian  symbol with unit power, and $n_0 $ is the additive white Gaussain noise at the user  with zero mean and variance $\sigma_n^2$.

To detect the transmit symbol, the user applies a one-tap equalizer $c$ and the estimate is given by $\hat{s} = cy$.
\subsection{CSI Uncertainty Model}
As in \cite{Zheng2009RobustBI}, the CSI errors are assumed to follow the complex Gaussian distribution, namely,
\begin{equation}
\begin{cases}
\textbf{G}  = \hat {\textbf{G}} + \Delta \textbf{G}, \\
\textbf{h}_d = \hat{\textbf h}_d + \Delta \textbf{h}_d, \\
\textbf{h}_r = \hat{\textbf h}_r + \Delta \textbf{h}_r,
\label{model:channel=estimate+error}
\end{cases}
\end{equation}
where $ \hat {\textbf G}$, {$\hat {\textbf h}_r$}, and $\hat{\textbf h}_d$ denote  the estimated CSI, and $\Delta \textbf{G}$, {$\Delta {\textbf h}_r$}, and $\Delta{\textbf h}_d$ are the corresponding CSI
errors whose entries are i.i.d. zero-mean complex  Gaussian  with variances of $\sigma_g^2$, $\sigma_r^2$, and $\sigma_d^2$, respectively.

\subsection{Problem Formulation}	
As in \cite{Xing2010RobustJD, Shen2014RobustTF}, the criteria of minimizing the average mean squared error (MSE) is adopted to ensure statistical robustness.
Hence, we first write the objective function in the form of  MSE averaged over all  CSI errors:
\newcounter{TempEqCnt}
\setcounter{TempEqCnt}{\value{equation}}
\setcounter{equation}{5}
\begin{figure*}[hb]
	\hrulefill
	\begin{align}
	\textbf{A} = \hat {\textbf G }^H \boldsymbol \Theta^H \hat {\textbf h}_r\hat {\textbf h}_r^H \boldsymbol \Theta \hat{\textbf G} + \hat{\textbf{h}}_d\hat {\textbf h}_r^H \boldsymbol \Theta \hat{\textbf G} + \hat {\textbf G }^H \boldsymbol \Theta^H \hat {\textbf h}_r\hat{\textbf{h}}_d^H +\hat{\textbf{h}}_d\hat{\textbf{h}}_d^H+ \sigma_g^2||\hat{\textbf h}_r||^2 \textbf I_M + \sigma_r^2 \hat {\textbf G }^H \hat {\textbf G } + (N\sigma_r^2\sigma_g^2 + \sigma_d^2)\textbf I_M	
	\label{eq:Alpha, bottom long equation}
	\end{align}			
\end{figure*}
\setcounter{equation}{\value{TempEqCnt}}
\begin{align}
e(\textbf w, c, \boldsymbol \Theta) &= \mathbb{E} \big \{|\hat{s} - s|^2\big \},
\label{eq:average MSE}
\end{align}
where the expectation is taken over the data symbol, additive Gaussian  noise, and the CSI errors. Substituting the CSI error model (\ref{model:channel=estimate+error}) into (\ref{eq:average MSE}) and using the fact that $\mathbb{E}\{\Delta \textbf{G}^H \textbf{H} \Delta \textbf{G}\} = \sigma_g^2{\rm Tr}\{\textbf{H}\}\textbf{I}_M$ for any $\textbf{H} \in \mathbb C^{N\times N}$, the MSE expression (\ref{eq:average MSE}) can be computed as
\begin{align}
e(\textbf w, c, \boldsymbol \Theta)
&=  |c|^2(\textbf w^H \textbf{A} \textbf{w} +\sigma_n^2) -  \textbf{w}^H\boldsymbol  \alpha c^*  - c\boldsymbol  \alpha^H\textbf{w} +1,  \label{eq:e_MSE_no_decoder_a}
\end{align}	
where $\textbf{A}$ is given by (\ref{eq:Alpha, bottom long equation}) at the bottom of the next page and $\boldsymbol  \alpha = (\hat{\textbf G}^H\boldsymbol \Theta^H \hat{\textbf h}_r+\hat{\textbf h}_d)$. Subject to the power constraint at  the AP and the unit-modulus constraint at the IRS, the joint design of  the transceiver and IRS phase shifts can be formulated as
\begin{align*}
&\min \limits_{\textbf w, c, \boldsymbol \Theta} \quad e(\textbf w, c, \boldsymbol \Theta) 	\tag{P1} \label{P1:MMSE}\\
& \begin{array}{r@{\quad}r@{}l@{\quad}l}
{\rm s.t.}&							
\begin{cases}
||\bm w||^2 \leq P_0, \\
0 \leq \theta_n < 2\pi, \quad \forall n = 1, \ldots, N.
\end{cases}\\
\end{array} .
\end{align*}
Note that the objective function (\ref{eq:e_MSE_no_decoder_a}) is non-convex with respect to (w.r.t.) \textbf{w}, $c,$ and $\boldsymbol{\Theta}$, which makes the optimization problem (\ref{P1:MMSE}) very difficult to solve.	
In the following, we propose an AO  method to solve it.

\section{Alternating Optimization}

In this section, we focus on solving problem (\ref{P1:MMSE}) via AO. 
Specifically, the transceiver  $\textbf{w}, c$,  and IRS phase shift matrix $\boldsymbol{\Theta} $  are optimized iteratively in an alternating manner until convergence. 

\subsection{Updating \{\textbf{w}, c\} Given $\Theta$}
First we update the beamfoming vector \textbf{w} and one-tap equalizer $c$ for a  given phase shift matrix $\boldsymbol{\Theta} $. Specifically, given  an arbitrary fixed $\boldsymbol{\Theta} $, the original problem (\ref{P1:MMSE}) can be reformulated as follows:
\setcounter{equation}{6}
\begin{align}
&\min \limits_{\textbf w, c } \quad |c|^2(\textbf{w}^H\textbf{A}\textbf{w} + \sigma_n^2) - \textbf{w}^H\boldsymbol  \alpha c^* - c\boldsymbol  \alpha^H\textbf{w}	 \label{P2:Updating w Given Theta}\\
&\;	{\rm s.t.} \qquad							
||\textbf{w}||^2 \leq P_0.  \nonumber
\end{align}
It can be observed that the objective function is non-convex w.r.t. \textbf{w} and $ c$,
hence, we  optimize $\textbf{w}$  and $c$ alternatingly. 
Given $\textbf{w}$, the optimal equalizer $c$ for the problem (\ref{P2:Updating w Given Theta}) is known to be the classical Wiener filter  \cite{Ingle2000StatisticalAA}:
\begin{equation}
c = \frac{\textbf{w}^H\boldsymbol{\alpha}}{\textbf{w}^H\textbf{A}\textbf{w}+\sigma_n^2}.	\label{eq:the closed-form solution c}
\end{equation}
Then, for a fixed $c$, the optimal beamforming vector \textbf{w} derived by using the Lagrangian method.  The Lagrangian function for (\ref{P2:Updating w Given Theta}) is given by
\begin{equation}
L(\textbf w, \lambda) = e(\textbf{w}) + \lambda(||\textbf w||^2 - P_0),
\label{eq:Lagrangian function}
\end{equation}
where $e(\textbf{w})$ is given in (\ref{P2:Updating w Given Theta}) and $\lambda$ is the Lagrange multiplier associated with $\textbf{w}$. Taking the derivative of ($\ref{eq:Lagrangian function}$)  w.r.t. beamformer $\textbf w^*$,
we can find the optimal solutions for (\ref{P2:Updating w Given Theta}) with the Karush–Kuhn–Tucker  conditions \cite{Boyd2004ConvexO}:
\begin{gather}
\textbf w = (|c|^2\textbf{A} + \lambda\textbf I_M )^{-1}\boldsymbol{\alpha}c^*
\label{eq:the closed-form solution w}\\
\lambda \ge 0 \tag{10a}\\
||\textbf w||^2 - P_0 \leq 0\tag{10b}\\
\lambda(||\textbf w||^2 - P_0) = 0\tag{10c} \label{condition:KKT-c}
\end{gather}
It is obvious that (\ref{eq:the closed-form solution w}) is sufficient and necessary for the optimal.
As shown in (\ref{eq:the closed-form solution w}), the optimal $\textbf{w}$ depends on the Lagrange multiplier $\lambda$,
thus $\lambda$ should be  calculated first before computing the beamformer $\textbf w$. According to (\ref{condition:KKT-c}), either $\lambda = 0$ or  $||\textbf w||^2 = P_0$ must hold.
Hence, if $\lambda = 0$ and   $||\textbf w||^2 - P_0 \leq 0$ is satisfied, then $\lambda = 0$. In contrast, if $\lambda = 0$, but $||\textbf w||^2 - P_0 \leq 0$ is not satisfied, then we have to solve the equation $||\textbf w||^2 = P_0$, which can be done numerically using the bisection search method.
After $\lambda$ is obtained, we then calculate the beamformer $\textbf{w}$ according to (\ref{eq:the closed-form solution w}).

\subsection{Updating $\Theta$ Given \{\textbf{w}, c\}}
Now we optimize the phase shift matrix  $\boldsymbol{\Theta}$  with fixed \textbf{w} and $c$. For simplicity, we omit the constant terms in (\ref{eq:e_MSE_no_decoder_a}) and rewrite the objective function of (\ref{P1:MMSE}) as
\begin{align}
f(\boldsymbol{\Theta}) &=  \big (\hat {\textbf h}_r^H \boldsymbol \Theta \hat{\textbf G} \textbf w\textbf w^H\hat {\textbf G }^H \boldsymbol \Theta^H \hat {\textbf h}_r +\hat{\textbf h}_d^H\textbf{w}\textbf{w}^H\hat {\textbf G }^H \boldsymbol \Theta^H \hat {\textbf h}_r\nonumber\\
&\quad+\hat {\textbf h}_r^H \boldsymbol \Theta \hat{\textbf G}\textbf{w}\textbf{w}^H\hat{\textbf{h}}_d\big)|c|^2  - 2\mathcal{R}(\hat{\textbf h}_r^H \boldsymbol \Theta \hat{\textbf G}\textbf w c). \label{eq:f_design_theta}
\end{align}
Let $\textbf{v} \triangleq [v_1,\cdots, v_N]^H$ where $v_i = e^{j\theta_i}$ for all $ i = 1, \cdots, N$, $\boldsymbol \Phi = {\rm diag}(\hat {\textbf h}_r^H) \hat{\textbf G} \textbf w c$ and  ${d} = \hat{\textbf h}_d^H\textbf{w}c$, then we  have  $ \hat {\textbf h}_r^H \boldsymbol \Theta_{i+1} \hat{\textbf G} \textbf w c = \textbf{v}^H \boldsymbol{\Phi} $. As such, (\ref{eq:f_design_theta}) can be rewritten as
\begin{align}
f(\textbf{v}) =   \textbf v^H \boldsymbol{\Phi} \boldsymbol{\Phi}^H \textbf{v} -2\mathcal{R}(\textbf v^H \boldsymbol{\Phi}(1-d^*)).  \label{eq:J_v_simple}
\end{align}
Hence, the corresponding optimization problem can be recast as nonconvex quadratically constrained quadratic programs (QCQPs)
\begin{align}
&\min \limits_{\textbf v} \quad \textbf v^H \boldsymbol{\Phi} \boldsymbol{\Phi}^H \textbf{v} - 2\mathcal{R}(\textbf v^H \boldsymbol{\Phi}(1-d^*)) 	 \label{P3:nonconvex QCQP}\\
&
\;	{\rm s.t.}						
\quad\ |v_n| = 1, \forall n = 1,\cdots, N. \nonumber
\end{align}
Due to the unit modulus constraint, the above problem is non-convex, and belongs to the class of NP-hard problems.
The conventional approach is to reformulate the above problem as semidefinite programming problem via matrix-lifting \cite{Zhang2006ComplexQO}.
However, as the number of reflecting elements grows large,  the implementation of the matrix-lifting procedure is challenging.

Therefore, we propose a MM based method to solve (\ref{P3:nonconvex QCQP}).  The MM algorithm is an iterative technique to find an absolute minimizer. Instead of minimizing $f(\textbf{v})$, this method minimizes a majorization function of $f(\textbf{v})$ at each iteration point. The $k$th  majorizer for the objective function should satisfy the following two conditions:
\begin{align}
g(\textbf{v}, \textbf{v}_{k-1}) \geq f(\textbf{v}), \forall \textbf v,  \nonumber \\
g(\textbf{v}_{k-1}, \textbf{v}_{k-1}) = f(\textbf{v}_{k-1}),  \label{condition:MM algorithm majorizer}
\end{align}
where $\textbf{v}_{k-1}$ is the value of $\textbf{v}$ at the $(k-1)$th iteration. Indeed, the function $g(\textbf{v}, \textbf{v}_{k-1})$ is an upper bound of the function $f(\textbf{v})$ and the equality is achieved  at point $\textbf{v}_{k-1}$. To ensure a monotonically decreasing
sequence of the function values, each iterative value $\textbf v_k$ follows the update rule  $\textbf{v}_{k} = \rm {argmin} \  g(\textbf{v}, \textbf{v}_{k-1})$. So we have:
\begin{align}
f(\textbf{v}_k) \leq g(\textbf{v}_k, \textbf{v}_{k-1}) \leq g(\textbf{v}_{k-1}, \textbf{v}_{k-1}) = f(\textbf{v}_{k-1}). \label{condition:MM algorithm decreasing poperty}
\end{align}
Hence, the key in MM algorithm is to determine the majorizer $g(\textbf{v}, \textbf{v}_{k-1})$ such that the majorized problem is easy to solve.  To apply the MM technique, we first rewrite  problem (\ref{P3:nonconvex QCQP}) as
\begin{align}
\qquad&\min \limits_{\textbf v} \quad \textbf v^H  \textbf Q\textbf{v} - 2\mathcal{R}(\textbf v^H \textbf{q}) 	 \label{P4:equivalent  problem P3}\\
\qquad&
\;{\rm	s.t.}					
\quad\ |v_n| = 1, \forall n = 1,\cdots, N, \nonumber
\end{align}
where $\textbf Q = \boldsymbol{\Phi} \boldsymbol{\Phi}^H$ is a positive semidefinite matrix and $\textbf{q} = \boldsymbol{\Phi}(1-d^*) $. Invoking the Claim 1 of \cite{Qiu2016PRIMEPR},   the function $\textbf v^H  \textbf Q  \textbf{v}$ can be majorized  by $\textbf v^H  \textbf H  \textbf{v} + 2\mathcal{R}(\textbf v^H  (\textbf Q - \textbf H)  \textbf{v}_0) + \textbf v_0^H  (\textbf H - \textbf Q)  \textbf{v}_0$ at every $\textbf v_0 \in \mathbb{C}^N$, where $\textbf H $ is a  fixed matrix such that $ \textbf H \succeq \textbf{Q}$. Thus, the majorized problem of (\ref{P4:equivalent  problem P3}) can be expressed as
\begin{align}
&\min \limits_{\textbf v } \quad \textbf v^H  \textbf H  \textbf{v} + 2\mathcal{R}(\textbf v^H  (\textbf Q - \textbf H)  \textbf{v}_0)  - 2\mathcal{R}(\textbf v^H \textbf{q})	 \label{P5:majorization problem}\\
& \;
{\rm s.t.}						
\quad\ |v_n| = 1, \forall n = 1,\cdots, N. \nonumber
\end{align}
Let $\textbf H = \lambda_{max}(\textbf Q)\textbf I $ where $\lambda_{max}(\textbf Q)$ is the largest eigenvalue of matrix \textbf{Q}, so that the first term of (\ref{P5:majorization problem}) is a constant. By discarding constant terms w.r.t. \textbf{v}, the new majorization problem at the ($k+1$)th iteration is
\begin{align}
\qquad&\min \limits_{\textbf v } \quad \mathcal{R}(\textbf v^H \textbf{u})	 \label{P6:new majorization problem}\\
& \;
{\rm s.t.}							
\quad\ |v_n| = 1, \forall n = 1,\cdots, N, \nonumber
\end{align}
where \textbf u = $ {(\textbf Q - \lambda_{max}(\textbf Q)\textbf I)}  \textbf{v}_k - \textbf{q}$ is  a constant w.r.t. the variable \textbf{v} since the vector $\textbf{v}_k$ is known
beforehand by generating  at iteration $k$.
Thus, the optimal solution to problem (\ref{P6:new majorization problem})  is given by
\begin{align}
\textbf{v}_{k+1}^* = -e^{j{\rm arg}(\textbf{u})}
\label{eq:closed-form solution v*}
\end{align}
at the ($k+1$)th iteration. Since the monotonicity of the MM algorithm ensures that $f(\textbf{v}_i) \leq f(\textbf{v}_j)$ for all $i > j$, \textcolor{black}{we can repeat the above steps to find a stationary point and the phase shifts  $\boldsymbol{\Theta}$ can be easily recovered from $\textbf{v}^*$.}

\subsection{Overall Algorithm Description}
{In summary, the overall AO algorithm yields a simple closed-form solution at every iteration, which is given in Algorithm \ref{alg:overall AO}. As shown, the optimal solutions \{$\textbf{w}^t, c^t$\}  and locally optimal solution $\boldsymbol{\Theta}^t$ are obtained alternatingly, with superscript $t$ denoting the $t$th iteration.}
\begin{algorithm}[H]
	\caption{Proposed AO Algorithm}
	\label{alg:overall AO}
	\begin{algorithmic}[1]
		\STATE {\textcolor{black}{Initialize  $\boldsymbol{\Theta}^1$ with random phases,  $\textbf{w}^1 = \frac{\sqrt{P_0}}{\sqrt{M}}\textbf{1}$,  and $\textbf{v}^1 = {\rm diag}(\boldsymbol{\Theta}^1)$. Set iteration number $t = 1$.}}   
		\REPEAT
		\STATE Update  $c^{t}$  by (\ref{eq:the closed-form solution c})  given $\boldsymbol{\Theta}^t$ and $\textbf{w}^t$.
		\STATE Update  $\textbf{w}^{t+1}$ by  (\ref{eq:the closed-form solution w})  given $\boldsymbol{\Theta}^t$ and $ c^t$.
		\STATE Optimize $\textbf{v}^{t+1}$ according to (\ref{eq:closed-form solution v*})  given  $\{\textbf{w}^{t+1}, c^t\}$ and update  $\boldsymbol{\Theta}^{t+1}$ from $\textbf{v}^{t+1}$.
		\UNTIL \textcolor{black}{the decrease of the MSE is below  $\epsilon > 0$.}
		
	\end{algorithmic}
\end{algorithm}
\subsection{Convergence and Complexity Analysis}
The proposed AO algorithm can be shown to converge as follows. 
Recall the objective function of problem (\ref{P1:MMSE}) and it follows that
\begin{align}
e(\textbf{w}^{t+1},c^{t+1},\boldsymbol{\Theta}^{t+1})&\leq e(\textbf{w}^{t+1},c^{t+1}, \boldsymbol{\Theta}^{t}) \nonumber\\
&\leq e(\textbf{w}^{t},c^{t+1}, \boldsymbol{\Theta}^t)\nonumber\\
&\leq e(\textbf{w}^{t},c^{t},\boldsymbol{\Theta}^t).
\label{eq:convergence analysis}
\end{align}
The first inequality holds due to the non-increasing property of the
general MM scheme. The last two inequalities come from (\ref{eq:the closed-form solution w}) and (\ref{eq:the closed-form solution c}), i.e., $\textbf{w}^{t+1}$ and  $c^{t+1}$ corresponds to the minimizer of  $e(\textbf{w},c^{t+1}, \boldsymbol{\Theta}^t)$ and $e(\textbf{w}^t,c, \boldsymbol{\Theta}^t)$, respectively.

Furthermore, we briefly discuss the computational complexity of the proposed AO algorithm. In each iteration, the  algorithm yields  simple closed-form solutions and only requires basic matrix operations. Firstly, as for  updating the transcievr $\textbf{w}, c$,  the complexity to calculate \textbf{A} is on the order of $\mathcal{O}(M^2N) $. Also,  the complexity to find the optimal $\lambda$ is about $\mathcal{O}(log(m)M^3)$, where $m$ denotes the interval length of the bisection search. Secondly, as for updating the IRS phase shifts, the  algorithm requires the computational complexity  of $\mathcal{O}(n(NM+N^2)) $ for the matrix multiplications, where $n$ represents the number of MM algorithm iterations.   The overall  computational complexity of each iteration is given by $\mathcal{O}(M^2N+log(m)M^3+n(NM+N^2))$.

\section{Simulation Results}

In this section, numerical simulations are conducted to evaluate the performance of the proposed system. We consider a schematic system  as shown in Fig.\ref{sys_model}  with $M = 4$ transmit antennas.  \textcolor{black}{ We assume that the locations of the AP and IRS are (0 m, 0 m) and (100 m, 0 m), while the user is located at (100 m, 20 m).
The large-scale path loss is modeled as $L(d) = L_0d^{-\alpha}$,
where $L_0$ is the path loss at the reference distance 1 m, $d$ is the link distance in meters and $\alpha$ is the corresponding path loss exponent.} In simulations, the path loss exponents $\alpha$ for channels with and without line of sight (LoS) components are respectively set to be 2 and 3.
Considering the existence of LoS components, the channels between the AP-IRS and the IRS-user are modeled as  Ricean fading with Ricean factor $K = 10$. Furthermore, the  direct channel  ${\textbf{h}}_d$ is assumed to be Rayleigh flat-fading.
We also set $\sigma_g^2 = \sigma_r^2 = \sigma_d^2 = \sigma^2$ for simplicity.
\textcolor{black}{The other parameters are set as follows: $L_0 = -30$ dB, $\epsilon = 10^{-4}, \epsilon_{mm} = 10^{-8}, $ and  $\sigma_n^2 = -110$ dBm.} For performance comparison, the following four schemes are considered:
1) The proposed robust design, which jointly optimizes the transceiver and IRS phase shifts;
2) The non-robust scheme, which optimizes the system as if $\hat{\textbf h}_r, \hat{\textbf{h}}_d$ and $ \hat{\textbf G}$ are perfect;
\textcolor{black}{3) The discrete phase shifts scheme, which  quantities the optimized continuous phase shifts to its nearest values.
}
4) The scheme when IRS is not deployed, which simply optimizes the transceiver 
with $\boldsymbol{\Theta} = \textbf{0}$.   

\begin{figure}[tb!]
	\centering 
	\includegraphics[scale=0.6]{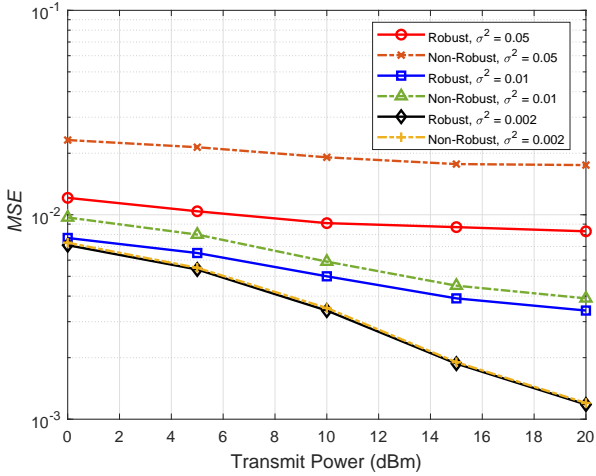}
	\caption{The MSE of Robust and Non-Robust versus transmit power where $N = 40$.}
	\label{figure:MM_SNRvsMSE}
\end{figure}

In Fig.\ref{figure:MM_SNRvsMSE}, we compare  the average MSE of the robust and non-robust schemes with  {$N = 40$}. Each point in Fig.\ref{figure:MM_SNRvsMSE} is an average over 1000 independent channel realizations.
As can be readily observed,  the proposed robust design method outperforms the conventional non-robust design scheme in all CSI error configurations. In addition, the performance gap depends on the accuracy of CSI, and the advantage of the proposed robust design method is most pronounced when the CSI error is large, i.e., $\sigma^2=0.05$. As the CSI error becomes smaller, the performance gap gradually diminishes. Finally, it is observed that the average MSE reduces as the transmit power increases, as expected.

\begin{figure}[tb!]
	\centering
	\includegraphics[scale=0.6]{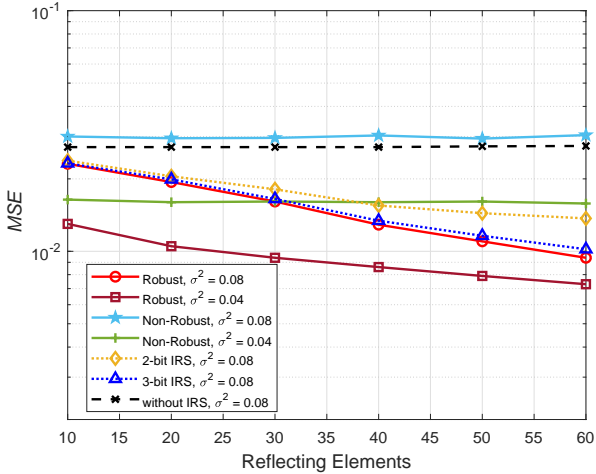}
	\caption{The MSE of the four schemes versus $N$ where $P_0$ = 10 dBm.}
	\label{figure:MM_NvsMSE}
\end{figure}

\textcolor{black}{
Fig.\ref{figure:MM_NvsMSE} shows the impact of the number of reflection elements on the average MSE of different schemes. First, we can see that the  performance of the robust design improves  consistently  with the increase of IRS elements, especially under large $\sigma^2$. 
Besides, it is observed that the finite phase resolution scheme suffers performance loss compared to IRS with continuous phase shifts as expected.
However as the number of quantization bits increases, for instance, with 3-bit phase shifter, the performance degradation rapidly becomes negligible.
Moreover, for the non-robust scheme, the MSE is almost unchanged with the increase of IRS elements. The reason is that, when $N$ becomes large, the aggregate CSI mismatch  also increases alongside the corresponding channel dimension.}
Finally, it also illustrates that the MSE performance of the case without deploying IRS is extremely poor, which explains the  significance of the IRS enhancement.

\section{Conclusions}
In this paper, we studied the joint design of transceiver and
phase shifts  for an IRS-aided MISO system with imperfect CSI. We proposed an alternating  beamforming optimization algorithm based on the Lagrangian method and MM technique. The proposed algorithm is shown to be robust against CSI errors, and achieves significantly better MSE performance compared to the non-robust design methods. For future research, it is promising to extend the robust design frame to the more general cases, such as MIMO or multi-user scenarios.



\end{document}